\documentclass[prl,twocolumn,showpacs,superscriptaddress,floatfix,amsmath,amssymb]{revtex4}

\usepackage{graphicx}
\usepackage{dcolumn}
\usepackage{amsfonts}
\usepackage{bm}

\begin{document}

\preprint{APS/123-QED}

\title{Magnetic excitations in \{Mo$_{72}$Fe$_{30}$\}}

\author{V. O. Garlea}
 \email{garleao@ornl.gov}
\affiliation{Ames Laboratory, Department of Physics and Astronomy,
Iowa State University, Ames, IA, 50011, USA}
\author{S. E. Nagler}
\affiliation{Oak Ridge National Laboratory, Oak Ridge, TN 37831 USA}
\author{J. L. Zarestky}
\affiliation{Ames Laboratory, Department of Physics and Astronomy,
Iowa State University, Ames, IA, 50011, USA}
\author{C. Stassis}
\affiliation{Ames Laboratory, Department of Physics and Astronomy,
Iowa State University, Ames, IA, 50011, USA}
\author{D. Vaknin}
\affiliation{Ames Laboratory, Department of Physics and Astronomy,
Iowa State University, Ames, IA, 50011, USA}
\author{P. K\"{o}gerler}
\affiliation{Ames Laboratory, Department of Physics and Astronomy,
Iowa State University, Ames, IA, 50011, USA}
\author{D. F. McMorrow}
\affiliation{Ris{\o} National Laboratory, DK-4000, Roskilde,
Denmark} \affiliation{Department of Physics and Astronomy,
University College London, UK}
\author{C. Niedermayer}
\affiliation{Laboratory for Neutron Scattering ETHZ \& PSI, CH-5232
Villigen PSI, Switzerland}
\author{D. A. Tennant}
\affiliation{School of Physics and Astronomy, University of St
Andrews, St Andrews, FIFE KY16 9SS, Scotland, UK}
\author{B. Lake}
\affiliation{Clarendon Laboratory, University of Oxford, Parks Road,
Oxford OX1 3PU, UK} \affiliation{Ames Laboratory, Department of
Physics and Astronomy, Iowa State University, Ames, IA, 50011, USA}
\author{Y. Qiu}
\affiliation{NIST Center for Neutron Research, Gaithersburg, MD
 \& U. Maryland, College Park, MD, USA}
\author{M. Exler}
\affiliation{Universit\"{a}t Osnabr\"{u}ck, Fachbereich Physik -
D-49069 Osnabr\"{u}ck, Germany}
\author{J. Schnack}
\affiliation{Universit\"{a}t Osnabr\"{u}ck, Fachbereich Physik -
D-49069 Osnabr\"{u}ck, Germany}
\author{M. Luban}
\affiliation{Ames Laboratory, Department of Physics and Astronomy,
Iowa State University, Ames, IA, 50011, USA}

\date{\today}

\begin{abstract}
We report cold-neutron inelastic neutron scattering measurements on
deuterated samples of the giant polyoxomolybdate magnetic molecule
\{Mo$_{72}$Fe$_{30}$\}. The 30 $s$ = 5/2 Fe$^{3+}$ ions occupy the
vertices of an icosidodecahedron, and interact via antiferromagnetic
nearest neighbor coupling. The measurements reveal a band of
magnetic excitations near E $\approx$ 0.6 meV. The spectrum broadens
and shifts to lower energy as the temperature is increased, and also
is strongly affected by magnetic fields. The results can be
interpreted within the context of an effective three-sublattice spin
Hamiltonian.
\end{abstract}

\pacs{75.25.+z, 75.50.Ee, 75.75.+a, 78.70.Nx}
\maketitle

Magnetic molecules are ideal prototypical systems for the study of
fundamental problems in magnetism on the nanoscale level~\cite{1}.
As a result, their properties have been the subject of many
theoretical and experimental
investigations~\cite{2,3,4,5,6,7,8,9,10,11,12}. Of paramount
importance is the determination of the magnetic excitation spectrum.
However, for a molecule with $N$ magnetic ions of spin $s$ the
calculation of the $(2s+1)^{N}$ eigenstates and their energies
quickly becomes impractical for increasing $N$ and $s$. Neutron
scattering is the most effective and direct technique for
determining the magnetic energy levels, and there have been studies
of excitations in several magnetic molecules containing up to 12
spins~\cite{2,3,4,5,6,7}.

In this Letter we report cold-neutron inelastic scattering results
obtained on one of the largest magnetic molecules yet synthesized:
the polyoxomolybdate cluster \{Mo$_{72}$Fe$_{30}$\}. The
crystallographic structure is described by the space group
$R\overline{3}$ with the lattice constants:~$a \approx$ 55.13 \AA,
and $c\approx$ 60.19 \AA~\cite{9}. The molecule contains 30
Fe$^{3+}$ ions ($s$ = 5/2) occupying the vertices of an
icosidodecahedron. The magnetic ions are interlinked by
Mo$_{6}$O$_{21}$ fragments acting as super-exchange pathways,
resulting in nearest neighbor antiferromagnetic exchange $J
\vec{\textit{\textbf S}_i} \cdot \vec{\textit{\textbf S}_j}$ and a
singlet spin ground state. As the icosidodecahedron consists of 20
corner-sharing triangles circumscribing 12 pentagons the spins are
frustrated and show properties similar to the antiferromagnetic
Kagom\'{e} lattice~\cite{13}. For $s$ = 5/2 it is reasonable to
consider classical spin vectors as a starting point~\cite{8},
leading to a picture at \textit{T} = 0 of three sublattices of 10
parallel spins each, with orientations defined by coplanar vectors
offset by 120$^{\circ}$ angles. As discussed below, many features of
the observed scattering can be interpreted in the context of a
solvable three-sublattice effective Hamiltonian substituted for the
intractable Heisenberg Hamiltonian~\cite{8,10}.

Most of the neutron scattering experiments were performed on
deuterated samples to minimize the attenuation and incoherent
scattering from the hydrogen atoms. Characterization of the
deuterated samples by infrared and Raman spectroscopy and X-ray
diffraction confirmed that their properties were consistent with
those of non-deuterated samples studied earlier~\cite{9}.

The neutron measurements used polycrystalline samples of
approximately 10 g sealed in copper holders under He atmosphere.
Preliminary characterization by powder diffraction was performed at
the HB1A instrument at HFIR. Upon cooling from room temperature the
diffraction patterns in both deuterated and non-deuterated samples
exhibited a remarkable increase in the background over the entire
measured range of scattering angle. The presence of this scattering
at large wavevectors is a clear indication that it is not magnetic
in origin. It can be understood as arising from quenched static
structural disorder, and is manifested as a very large zero energy
peak in the inelastic scattering spectra.  It presents a significant
experimental challenge, and prevents a clean observation of low
energy magnetic scattering.

The inelastic neutron spectra were collected at low temperatures
using three different spectrometers: RITA2 at PSI, DCS at NIST,
OSIRIS at ISIS. The data presented here were obtained at OSIRIS
using a fixed final neutron energy E$_{f}$ = 1.845 meV. The results
obtained with DCS and RITA2 are consistent with those of OSIRIS. A
complete description of all of the results will be presented
elsewhere~\cite{14}.

\begin{figure}[tbp!]
\includegraphics{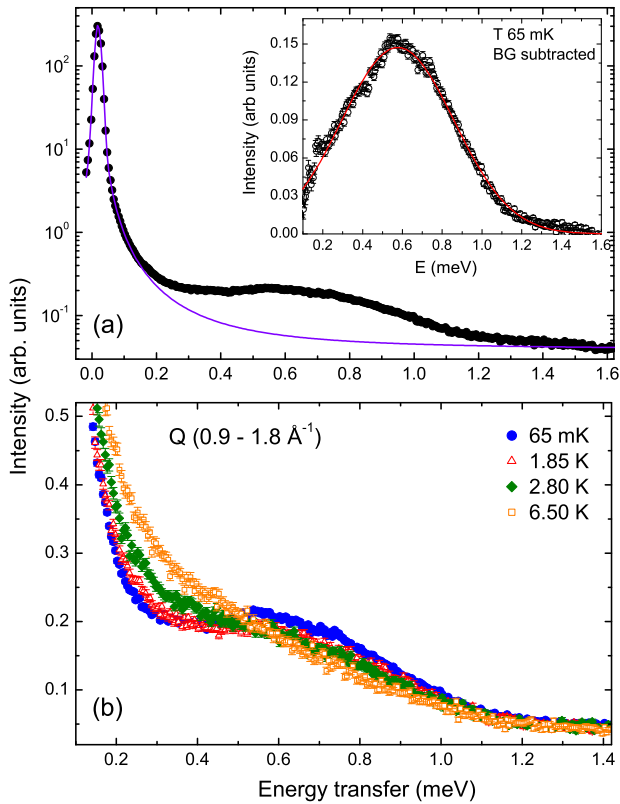}
\caption{\label{fig1} (a) Inelastic neutron scattering spectrum at a
nominal temperature of 65 mK. The intensity axis is on a logarithmic
scale. The solid line shows a best estimate of the non-magnetic
background (see text). Inset: The background subtracted scattering.
(b) Raw data for several different temperatures plotted on a linear
scale.}
\end{figure}

Typical data obtained are shown in Fig.~\ref{fig1}. The scans
plotted in all figures are integrated over the range Q = 0.9 to 1.8
\AA$^{-1}$. Positive energies correspond to neutron energy loss. The
upper panel (Fig.~\ref{fig1}(a)) shows a semi-log plot of the
spectrum at the base temperature, nominally \textit{T}= 65 mK. The
large elastic peak has a FWHM of 0.021(1) meV consistent with the
instrumental energy resolution. A much broader peak is visible as a
shoulder in the plot. The lower panel (Fig.~\ref{fig1}(b)) shows the
scattering on a linear scale at four different temperatures ranging
up to 6.5 K. At base temperature a clear peak is evident near an
energy transfer of 0.6 meV. As the temperature is raised this peak
weakens, and there is a notable increase in the intensity of the
scattering at low energies. This behavior of the intensity is
expected if the peak at 0.6 meV is due to a magnetic transition from
the ground state. The temperature-enhanced low energy signal could
in principle arise from lattice vibrations or from magnetic
scattering, and differentiating between these possibilities requires
careful analysis.

Cooling powders to mK temperatures is known to be extremely
difficult, and since the temperature is measured external to the
sample it is desirable to have an internal consistency check. This
can be accomplished in principle by checking the detailed balance
condition, but ambiguities in the scattering background, limited
data for neutron energy gain, and a resolution dependent on the
energy transfer limit the precision that can be achieved. Going
through this exercise suggests that the nominal sample temperatures
at 1.8 K and above are probably reliable, but the base temperature
to which the powder is actually cooled is less certain. The
evolution of the scattering shows that it is clearly well below 1.8
K but it is likely warmer than the nominal value 65 mK. However, our
data analysis does not depend on the precise value of the base
temperature.

To extract more quantitative information the base temperature
scattering was fit to a simple model. It was assumed that the
instrumental resolution at the elastic position can be described by
the sum of two co-centered peaks: a dominant Gaussian and a
Lorentzian with the latter accounting for the extended tails. The
nonmagnetic background scattering is taken as the sum of an elastic
term proportional to the resolution function plus an energy
independent constant. The magnetic scattering was assumed to be
represented by a single peak, with a Gaussian found to provide a
better fit than a Lorentzian. Least squares fitting of this simple
model to the data was carried out initially using the entire data
set. Alternatively, the analysis was repeated by determining the
model background alone by fitting to the data excluding a range in
the vicinity of the inelastic peak.  Subsequently a single magnetic
peak was fitted to the data with the above determined background
subtracted.  The results were found to be independent of the size of
the excluded range and consistent with the fit over the whole range.
The solid curve in Fig.~\ref{fig1}(a) illustrates the background,
and the inset shows the background subtracted data fit to a single
Gaussian with peak position 0.56(1) meV and FWHM 0.66(1) meV. The
numbers in parentheses are the estimated uncertainties in the last
digit, and account for both fitting errors and systematic effects
arising from different background estimations. As seen in the inset
to Fig.~\ref{fig1}(a) the overall distribution is modeled well by a
single peak, however the data shows additional structure, including
one or more shoulders on the low energy side. The observed energy
width of the shoulders and the overall distribution are intrinsic
since the resolution is roughly FWHM 0.02 meV. The signal for energy
transfers greater than 0.2 meV is independent of the assumptions
used in determining the background.

\begin{figure}[tbp!]
\includegraphics{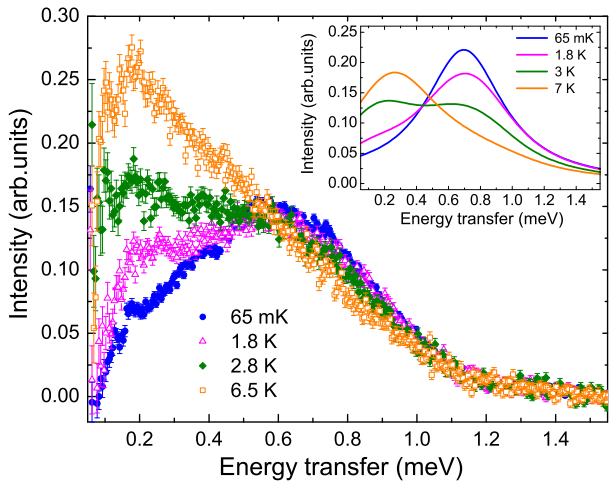}
\caption{\label{fig2} Temperature dependence of the inelastic
neutron scattering, plotted with the nonmagnetic background
subtracted. Inset: Theoretical scattering for several temperatures,
calculated using the quantum rotational band model with simplifying
assumptions (see text).}
\end{figure}

Further analysis was carried out to gain more insight into the
temperature dependence of the scattering. A thorough attempt was
made to fit the data for energy transfers less than 0.5 meV and
temperatures up to 6.5 K to models of both single and multiple
phonon scattering from solids or fluids~\cite{15}. This attempt was
unsuccessful. Subsequently, the data was examined in a
model-independent fashion, integrating the energy cuts over
different wavevector ranges varying from Q$_{L}$ = 1.0 -- 1.2
\AA$^{-1}$ to Q$_{U}$ = 1.6 -- 1.8 \AA$^{-1}$. It was verified that
for all temperatures up to 6.5 K there is no Q dependence in the
scattering to within the sensitivity of the measurement. This is a
significant result since the intensity for single phonon scattering
is expected to increase proportional to Q$^{2}$, and that of
multiphonon scattering should increase even more rapidly with
Q~\cite{15}. The intensity of phonon scattering in the range Q$_{U}$
should be stronger by a factor of 2.5 or more than that in Q$_{L}$,
contrary to the experimental observation. It can be concluded that
phonon scattering is insignificant over the wavevector and
temperature range of the data considered here.

Given this result, it is reasonable to assume that the nonmagnetic
background is temperature independent.  The background subtracted
scattering for several temperatures up to 6.5 K is shown in
Fig.~\ref{fig2}. With the uncertainties in temperature and
instrumental resolution there may be some amount of magnetic
scattering included as background at low energy transfers, possibly
contributing to an over-subtraction at energy transfers below 0.2
meV. Notwithstanding these ambiguities we believe that this data
provides a fair representation of the inelastic magnetic scattering
in \{Mo$_{72}$Fe$_{30}$\}.

\begin{figure}[tbp!]
\includegraphics{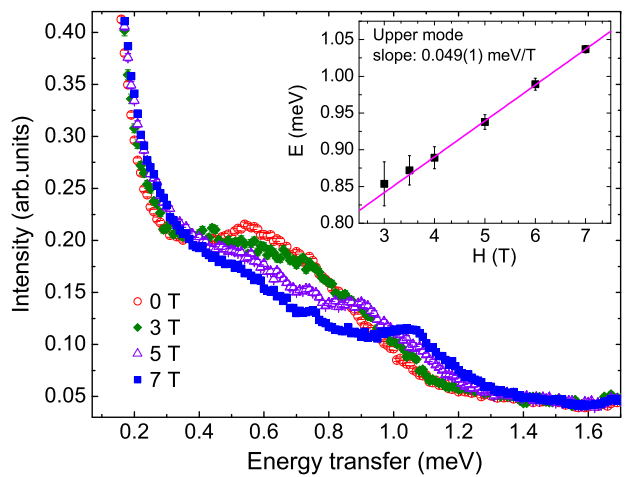}
\caption{\label{fig3} Scattering at base temperature for several
different values of applied magnetic field. Inset: Upper mode energy
as a function of field.}
\end{figure}

The magnetic nature of the observed spectrum was confirmed by
studying the effect of an externally applied magnetic field. As
illustrated by the base temperature data in Fig.~\ref{fig3} the
application of a field leads to a noticeable broadening of the peak.
With increasing magnetic field the intensity of the main peak near
0.6 meV decreases and the scattering distribution broadens.
Examining the difference between scattering at zero and non-zero
fields~\cite{14} shows that the spectrum in a field consists of a
central peak with two almost symmetric side bands, the positions of
which vary roughly linearly with field strength. The position of the
central peak shifts slightly with the applied field; the value at 7
T is estimated as 0.50(1) meV. Following the detailed field
dependence of the lower sideband quantitatively is difficult because
of the large background at low energy transfers. The higher energy
sideband can be observed more cleanly, and the peak position as a
function of field is plotted in the inset to Fig.~\ref{fig3}. For
very small fields the side peaks cannot be resolved, but over the
visible range the slope of the upper peak is 0.049(1) meV/T.

At present there is no rigorous theoretical calculation available
for a detailed comparison with the results of this experiment. The
large number of $s$ = 5/2 magnetic ions per molecule precludes the
diagonalization of the quantum Heisenberg Hamiltonian. However, the
main features of the energy spectrum can be established using a
solvable effective 3-sublattice Hamiltonian~\cite{8,10} as a
starting point. Defining the total spin on each sublattice as $
\tilde {\textit {\textbf S}_\alpha}$ the Hamiltonian in the absence
an external magnetic field is taken as $\frac{J}{5}\left( {\tilde
{\textit{\textbf S}_A} \cdot \tilde {\textit{\textbf S}_B} + \tilde
{\textit{\textbf S}_B} \cdot \tilde {\textit{\textbf S}_C} + \tilde
{\textit{\textbf S}_C} \cdot \tilde {\textit{\textbf S}_A} }
\right)$. This neglects anisotropy, which is estimated to be small
compared to $J$~\cite{11}. The resulting spectrum has a hierarchy of
discrete energy levels for each value of the total spin quantum
number $S$ given by
$E(S)=\frac{J}{10}\left(S(S+1)-S_{A}(S_{A}+1)-S_{B}(S_{B}+1)-S_{C}(S_{C}+1)
\right)$, where the spin quantum numbers span the range 0 $\leq S
\leq$ 75 and 0 $\leq S_{A,B,C}\leq$ 25. The low-lying excitations
form a sequence of  well-separated, highly degenerate ``rotational
bands'' with excitation energies depending quadratically on $S$. The
energy levels in the lowest two rotational bands are given as $E_1
\left( S \right)=\frac{J}{10}S\left({S+1} \right)$ and
$E_2\left(S\right)=5J+E_1\left(S\right)$ with degeneracies for $S
\leq 50$ D$_1 = (2S+1)^2$ and D$_2 = 27(2S+1)^2$. The corresponding
levels of the exact Heisenberg model can be  expected to differ by
splittings of these levels.

Using the value of $J$ = 1.57 K found to describe the susceptibility
data~\cite{9} and the selection rules $\Delta S, \Delta M = 0,\pm1$
for neutron scattering leads to an estimate of the gap between the
the ground state and the first excited state within the band
$E_{1}(S)$ as $J/5\approx$ 27 $\mu$eV. The large elastic background
precludes an observation of this mode in the present experiment.
Using the same energy scale, transitions from the ground state to
the second rotational band should produce a pronounced intensity
near 5$J\approx$ 0.67 meV, consistent with the broad band of
excitations seen at base temperature, as shown in the inset to
Fig.~\ref{fig1}.

The detailed lineshape of the base temperature scattering may
reflect the richer structure of the energy spectrum of the
nearest-neighbor Heisenberg model, as compared to that of the
rotational band model. It may also reflect the effects of weak
anisotropy. A recent approximate spin-wave calculation~\cite{12} for
the nearest-neighbor model with anisotropy predicts that several
modes should be visible in the region of interest. A full
understanding of the line shape requires further theoretical work.

The temperature dependence of the scattering as shown in
Fig.~\ref{fig2} can also be considered within the context of the
rotational band model. As the temperature increases above \textit{T}
$\geq J/5\approx $ 0.3 K, energy levels within the lowest band
become more populated, giving rise to the broadening and shifting of
the main excitation towards E = 0. A detailed comparison with
experiment requires an evaluation of matrix elements. A calculation
was carried out starting from standard formulas for magnetic neutron
scattering~\cite{15}, using a common nonzero matrix element for all
allowed transitions~\cite{16}. Thermal occupation of levels was
accounted for by a Boltzmann factor, and Dirac delta function
factors associated with allowed transitions were replaced by a
Lorentzian with a width of 0.3 meV. The inset to Fig.~\ref{fig2}
shows the resulting calculated curves using $J$ = 1.57 K. Despite
the simplified nature of the approximations made there is a striking
resemblance between the curves and the data. The peak position of
the theoretical curve at 65 mK is in reasonable agreement with
observations. For successively higher temperatures the peak broadens
and shifts to lower energies. Near 7 K the peak in the intensity
occurs at around 0.2 meV, similar to that seen experimentally. The
simplified rotational band model provides a reasonable explanation
for the qualitative behavior with temperature.

An extension of the rotational-band model to non-zero fields
predicts that the ground state gradually shifts to $S >$ 0 states
preserving the quadratic dependence on $S$ in both the lowest and
first rotational bands~\cite{10}. Zeeman splitting of the levels
leads to excitations whose energies vary linearly with the magnetic
field. However, a quantitative theoretical explanation of the field
dependence of the neutron scattering cross-section remains as an
open problem.

In summary, using inelastic neutron scattering we have measured the
temperature and magnetic field dependence of magnetic excitations in
the Keplerate molecular magnet \{Mo$_{72}$Fe$_{30}$\}. A solvable
three-sublattice model~\cite{10} accounts for the overall energy
scale and qualitative temperature dependence of the observed
inelastic scattering. The principal mode observed can be understood
as arising from transitions between the two lowest rotational bands.
A quantitative understanding of the detailed \textit{T} = 0
lineshape and the behavior in a magnetic field will require a more
sophisticated theory. We hope that these results will stimulate the
continued development of theoretical methods incorporating the
essential qualitative features and symmetries of the system, and
that these will prove useful for other systems where diagonalization
of the Hamiltonian cannot be performed. Finally we note that the
neutron scattering experiments on large magnetic molecules are
currently very difficult, nonetheless important new information is
attainable now, and next generation instrumentation presents
significant new opportunities.

The authors thank R. J. McQueeney and B. Normand for helpful
discussions. Work at ORNL is supported by the U.S. DOE under
Contract No. DE-AC05-00OR22725 with UT-Battelle, LLC and work at
Ames Laboratory was supported under Contract No. W-7405-Eng-82. The
work at NIST was supported in part by the NSF under Agreement No.
DMR-0086210.  The work at Universit\"{a}t Osnabr\"{u}ck was
supported by DFG No. SCHN-615/5-2. The OSIRIS measurements (DAT, BL)
were supported by UK EPSRC GR/N35038/01.

\end{document}